\documentclass[amsmath, amssymb, aps, twocolumn]{revtex4-1}
\usepackage{graphicx}
\usepackage{amsmath}
\usepackage{bm}
\usepackage{mathrsfs}
\usepackage{booktabs}

\begin{document}

\title{Phonon anomaly and anisotropic superconducting gap in non-centrosymmetric Li$_2$(Pd$_{1-x}$Pt$_x$)$_3$B}

\author{G. Eguchi$^1$}
\email{geguchi@scphys.kyoto-u.ac.jp}
\author{D. C. Peets$^1$} 
\altaffiliation[Current address: ]{Max Planck Institute for Solid State Research, D-70569 Stuttgart, Germany}
\author{M. Kriener$^1$}
\altaffiliation[Current address: ]{Institute of Scientific and Industrial Research, Osaka University, Osaka 567-0047, Japan}
\author{S. Yonezawa$^1$} \author{G. Bao$^2$} \author{S. Harada$^2$} \author{Y. Inada$^{2,3}$} \author{G.-q. Zheng$^2$} \author{Y. Maeno$^1$}
\affiliation{$^1$Department of Physics, Graduate School of Science, Kyoto University, Kyoto 606-8502, Japan}

\affiliation{$^2$Department of Physics, Okayama University, Okayama 700-8530, Japan}
\affiliation{$^3$Graduate School of Education, Okayama University, Okayama 700-8530, Japan}

\date{\today}

\begin{abstract}
We report the  systematic investigation of the specific heat of the noncentrosymmetric superconductor Li$_2$(Pd$_{1-x}$Pt$_x$)$_3$B as a function of $x$. There is a large deviation of the phononic specific heat from the conventional Debye specific heat for Pt-rich samples. In contrast with the fully-gapped conventional behavior for small $x$, a  power-law temperature dependence of the electronic specific heat is observed even at $x=0.5$. This results manifest a strongly-anisotropic or nodal superconducting gap even at $x=0.5$ and a nodal superconducting gap for $x \gtrsim 0.9$.

\begin{description}
\item[PACS numbers]
65.40.Ba, 74.20.Mn, 74.20.Rp, 74.25.Bt
\end{description}

\end{abstract}

\maketitle


Superconductivity in the absence of inversion symmetry has been attracting much interest because of possible novel phenomena such as the enhancement of the upper critical field $H_{\rm{c2}}$ beyond the Pauli limiting field, and the emergence of a topological superconducting state~\cite{Bauer2004PRL,Kimura2007PRL,Sigrist2006JMMM,Agterberg2007PRB,Fu2008PRL,Sato2009PRL,Sato2010PRB}. Mixing of spin-singlet and triplet pairings due to parity mixing is a key ingredient for such unconventional phenomena. Li$_2$(Pd$_{1-x}$Pt$_x$)$_3$B is one such superconductor exhibiting a novel superconducting state for $x > 0.8$, attributable to the dominance of spin-triplet pairing~\cite{Yuan2006PRL,Nishiyama2007PRL,Harada2012PRB}. This system offers an ideal stage for studying mixed-parity superconductivity, because of its non-magnetic and weakly-correlated nature~\cite{Nishiyama2007PRL,Lee2005PRB,Chandra2005116,Yoshida2008JPSJ,Tsuda2009JPSJ}.  
However, details of the superconducting gap structure and its evolution with $x$ are still unclear: several studies indicate a nodal superconducting gap for higher $x$~\cite{Yuan2006PRL,Nishiyama2007PRL,Takeya2007PRB,Harada2012PRB}, whereas others report fully-gapped superconductivity for all $x$~\cite{Hafliger2009JSNM}. To resolve this issue, systematic investigations of the $x$ dependence of physical properties are required.

In this paper, we report a systematic specific-heat investigation of Li$_2$(Pd$_{1-x}$Pt$_x$)$_3$B as a function of $x$, finding presence of phonon anomaly reflecting the lattice distortion~\cite{Harada2012PRB}. Power-law temperature dependence of the electronic specific heat is observed for $x\geq 0.5$. Considering a recent NMR study~\cite{Harada2012PRB}, our results suggest the development of a strongly-anisotropic superconducting gap even at $x=0.5$ and a nodal superconducting gap for $x \gtrsim 0.9$.


\begin{figure}
\begin{center}
\includegraphics[width=\columnwidth,clip]{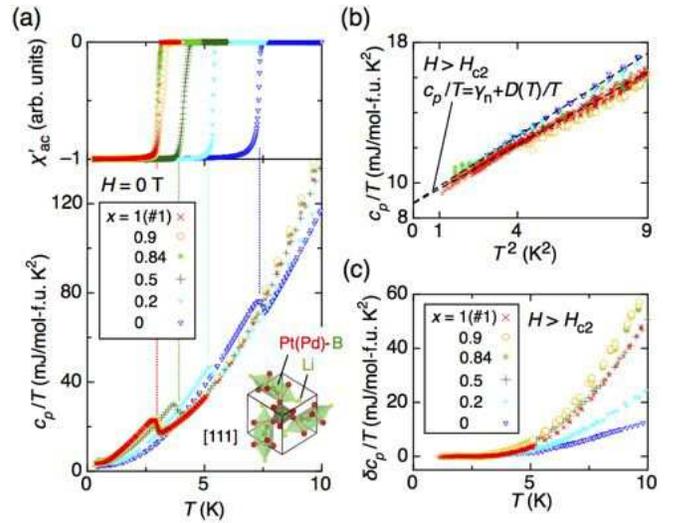}
\caption{(Color online) (a)~Temperature dependence of the ac susceptibility $\chi'_{\rm{ac}}$ and specific heat $c_p/T$ of Li$_2$(Pd$_{1-x}$Pt$_x$)$_3$B. The dotted vertical lines indicate the thermodynamic superconducting transition temperature $T_{\rm{c}}$. The inset shows the crystal structure seen from the [111] direction. (b)~Normal-state specific heat below 3~K in magnetic field $H$ larger than the superconducting upper critical field $H_{\rm{c2}}$~\cite{Darren2011PRB}. As examples, fits to the conventional Debye-Sommerfeld model $c_p/T=\gamma_{\rm{n}}+D(T)/T$ (see text) are shown for $x=0$ and $1$. (c)~Deviation of $c_p/T$ from the Debye-Sommerfeld model $\delta c_p/T=c_p/T-\gamma_{\rm{n}}-D(\varTheta_{\rm{D}}/T)/T$ for each sample.}
\label{cp}
\end{center}
\end{figure}

The polycrystalline samples used in this study were synthesized by two-step arc melting~\cite{Badica2005JPSJ}, and are from the same batches as the samples used in Ref.~\cite{Darren2011PRB}. The $x$ values refer to the nominal composition of Pt. The specific heat $c_p$ was measured down to 0.35 ~K with a commercial $^3$He refrigerator with a calorimeter (Quantum Design, PPMS).

\begin{figure}
\begin{center}
\includegraphics[width=\columnwidth,clip]{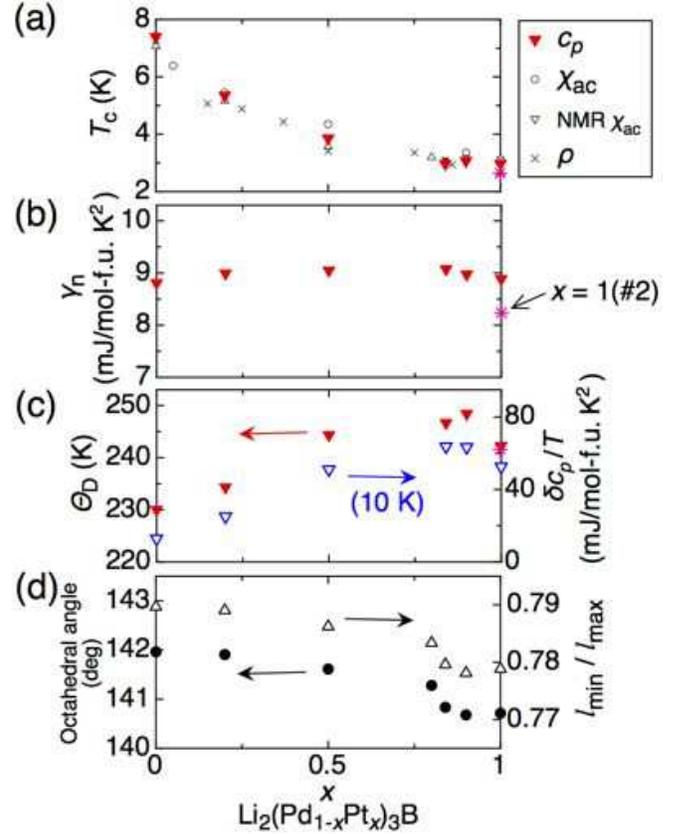}
\caption{(Color online)~Dependence on $x$ of (a) the superconducting transition temperature $T_{\rm{c}}$, determined from the specific heat $c_p$, from the ac susceptometry at low frequency $\chi_{\rm{ac}}$~\cite{Darren2011PRB} as well as at high frequency using an NMR coil~\cite{Harada2010PhysC}, and from the resistivity $\rho$~\cite{Shamsuzzamen2010JPCO}; (b)~the Sommerfeld coefficient $\gamma_{\rm{n}}$; (c)~the Debye temperature $\varTheta_{\rm{D}}$ and the deviation of $c_p/T$ from the Debye-Sommerfeld model at 10~K (see Fig.~\ref{cp}(c)); (d)~the $M$-$M$-$M$ bond angle between neighboring $M_6$B ($M=$Pd, Pt) octahedra and the ratio of the $M$-$M$ bond lengths, indicating an abrupt change in the octahedral distortion reproduced from Ref.~\cite{Harada2012PRB}.}
\label{param}
\end{center}
\end{figure}

The temperature dependence of the molar $c_p/T$ per formula unit (f.u.) of Li$_2$(Pd$_{1-x}$Pt$_x$)$_3$B is presented in Fig.~\ref{cp}(a), together with the ac susceptibility $\chi'_{\rm{ac}}$~\cite{Darren2011PRB}. For $x=1$, the two samples reported in this work are identified as $\#1$ and $\#2$. As indicated by the vertical dotted lines, the thermodynamic superconducting transition temperature $T_{\rm{c}}$ determined by entropy conservation agrees well with the temperature previously determined by ac susceptometry~\cite{Darren2011PRB}. The normal-state $c_p/T$ below 3~K in magnetic fields above $H_{\rm{c2}}$ is presented in Fig.~\ref{cp}(b). 
We find that the normal-state $c_p/T$ is independent of magnetic field (not shown), confirming the results in Ref.~\cite{Takeya2007PRB}. The data are fit with the conventional Debye-Sommerfeld model: $c_p/T=\gamma_{\rm{n}}+D(\varTheta_{\rm{D}}/T)/T$. Here, $\gamma_{\rm{n}}$ is the electronic specific heat coefficient (Sommerfeld coefficient) and $D(\varTheta_{\rm{D}}/T)$ is the Debye specific heat. The latter is defined as 
\begin{equation}
\begin{split}
D(\varTheta_{\rm{D}}/T)& =9N_{\rm{f.u.}}N_{\rm{A}}k_{\rm{B}} 
\bigg(\frac{T}{\varTheta_{\rm{D}}}\bigg)^3\int^{\varTheta_{\rm{D}}/T}_0 \frac{t^4 e^t}{(e^t-1)^2} dt, \notag
\end{split}
\end{equation}
where $\varTheta_{\rm{D}}$ is the Debye temperature, $N_{\rm{f.u.}}=6$ is the number of atoms per formula unit, $N_{\rm{A}}$ is Avogadro's number, and $k_{\rm{B}}$ is the Boltzmann constant. Here we neglect the difference between the constant-pressure and constant-volume specific heats. 
For all samples, the $c_p/T$ below 3~K is well fit by the model; as examples, we present the fitting results for $x=0$ and $x=1(\#1)$ with broken curves in Fig.~\ref{cp}(b). The fitting values of the two parameters are $\gamma_{\rm{n}}=8.8$~mJ/mol-f.u.\hspace{1pt}K$^2$ and $\varTheta_{\rm{D}}=230$~K for $x=0$, and $\gamma_{\rm{n}}=8.9$~mJ/mol-f.u.\hspace{1pt}K$^2$ and $\varTheta_{\rm{D}}=242$~K for $x=1(\#1)$. The variation of the parameters with $x$ is shown in Figs.~\ref{param}(b) and (c). Figure~\ref{cp}(c) shows $\delta c_p/T=c_p/T-\gamma_{\rm{n}}-D(\varTheta_{\rm{D}}/T)/T$ up to 10~K. Interestingly, at elevated temperatures, the $\delta c_p/T$ value increases significantly as $x$ increases, but slightly decreases at $x = 1$. This indicates that the low-energy phonon dispersion changes significantly with substitution. A large deviation from the conventional Debye-Sommerfeld model is evident for Pt-rich samples.

Now, we examine the $x$ dependences of $T_{\rm{c}}$ and the normal-state specific heat parameters. The $x$ dependence of the thermodynamic $T_{\rm{c}}$ in Fig.~\ref{param}(a) is wholly consistent with those in previous reports~\cite{Darren2011PRB,Harada2010PhysC,Shamsuzzamen2010JPCO}. In contrast to the significant change in $T_{\rm{c}}$, $\gamma_{\rm{n}}$ is nearly constant; effects other than the density of states influence $T_{\rm{c}}$.
The anomalous phononic specific heat $\delta c_p/T$, plotted in Fig.~\ref{param}(c), exhibits a steep increase from $x=0$ to the Pt-rich side, and forms a broad maximum around $x \sim 0.9$; similar tendency is seen in $\varTheta_{\rm{D}}$. Recently, it was reported that parameters characterizing the local lattice distortion, such as the $M$-$M$-$M$ bond angle between neighboring $M_6$B ($M=$Pd, Pt) octahedra and the ratio of the $M$-$M$ bond lengths, exhibit abrupt changes in the range $0.8 < x < 0.9$ as shown in Fig.~\ref{param}(d)~\cite{Harada2012PRB}. This indicates a lattice instability in this $x$ range. The observed enhancement of $\delta c_p/T$ is clearly related to this lattice instability, which should cause a phonon softening at $x \sim 0.9$. Note that $\delta c_p/T$ remains large even at $x =1$, indicating that this phonon mode remains soft in the Pt-end member.

\begin{figure}
\begin{center}
\includegraphics[width=\columnwidth,clip]{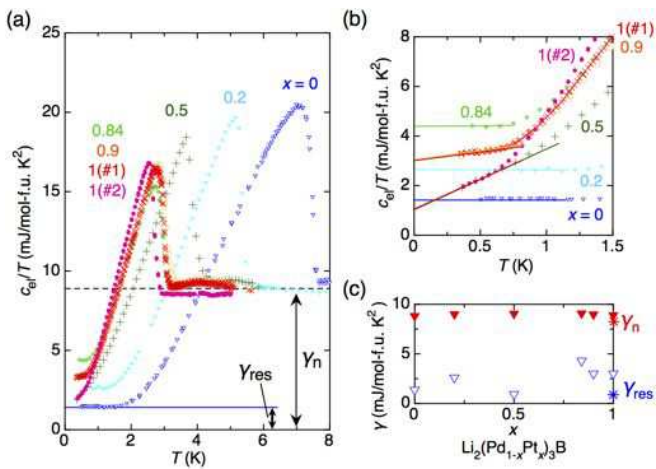}
\caption{(Color online)(a)~Temperature dependence of the electronic specific heat $c_{\rm{el}}/T$ for several Pt concentrations. For $x=1$, results for another sample labeled as $x=1$($\#2$) are also presented. The horizontal broken line represents the obtained Sommerfeld coefficient $\gamma_{\rm{n}}$ of the $x=0$ sample. The horizontal solid line represents the residual term $\gamma_{\rm{res}}$ of the $x=0$ sample. (b)~Enlarged view of low-temperature $c_{\rm{el}}/T$. Linear extrapolations of the low-temperature $c_{\rm{el}}/T$ so as to determine $\gamma_{\rm{res}}$ in each sample are shown with solid lines. (c)~Dependence of $\gamma_{\rm{n}}$ and $\gamma_{\rm{res}}$ on $x$.}
\label{cel}
\end{center}
\end{figure}

Next, we focus on the electronic specific heat $c_{\rm{el}}$ in the superconducting state, shown in Fig.~\ref{cel}(a). 
Since the contribution of the Debye phononic specific heat has been determined by the previous analyses, the zero-field $c_{\rm{el}}/T$ in the entire temperature range is deduced by subtracting $D(\Theta_{\rm{D}}/T)/T$ and a polynomial fit of $\delta c_p/T$ from the experimental result. For $x=1$, the result of another sample from a different batch ($\#2$) is also presented. As shown in the figure, $x=1$($\#2$) exhibits a smaller $c_{\rm{el}}/T$ values than $x=1$($\#1$) below 1~K. We indicate $\gamma_{\rm{n}}$ and its residual term $\gamma_{\rm{res}}$ for the $x=0$ sample in Fig.~\ref{cel}(a) as an example. Here we evaluate $\gamma_{\rm{res}} \equiv c_{\rm{el}}/T$~($T \rightarrow 0$~K) using a liner extrapolation.
The low-temperature $c_{\rm{el}}/T$ and its linear extrapolation for each sample is shown in Fig.~\ref{cel}(b). 

It is well known that the low-temperature $c_{\rm{el}}/T$ reflects the superconducting gap anisotropy. For $x=0$ and $0.2$, $c_{\rm{el}}/T$ does not depend on temperature below $\sim 1$~K. This indicates that the superconductivity for $x \leq 0.2$ is fully-gapped. On the contrary, the $c_{\rm{el}}/T$ for $x = 0.5, 0.9$ and $1$ exhibits noticeable temperature dependence. The samples with $x=0.5$ and $1$($\#2$) have small value of $\gamma_{\rm{res}}$ and thus should represent nearly intrinsic temperature dependence. In contrast, $x=0.9$ and $1$($\#1$) have higher $\gamma_{\rm{res}}$ with weaker temperature dependence. For $x=0.84$, $\gamma_{\rm{res}}$ is even higher and the temperature dependence diminishes at low temperatures. It has been known that the temperature dependence of $c_{\rm{el}}/T$ for $T \ll T_{\rm{c}}$ becomes weaker if impurity scatterings exist~\cite{Chang2000PRB}. Thus, the observed overall tendency is most consistent with an anisotropic or nodal superconducting gap for $x \gtrsim 0.5$. These results are consistent with the specific heat results in Ref.~\cite{Takeya2007PRB}. We also compare these results to the NMR spin-lattice relaxation rate $1/T_1$ in Ref.~\cite{Harada2012PRB}. The temperature dependence of $1/T_1$ taken down to $\sim 1.5$~K for $x=0.5$ and $0.8$ is fit using conventional BCS theory. Thus, it is important to examine the behavior of $1/T_1$ at lower temperatures in order to resolve the apparent inconsistency.

The cation-substitution-dependence of $\gamma_{\rm{res}}$ displayed 
in Fig.~\ref{cel}(c) shows no systematic trends. This is consistent with the interpretation that $\gamma_{\rm{res}}$ originates from defects or impurities. Further discussion is based on the $x=0$, $0.2$, $0.5$, and $1$($\#2$) samples, which exhibit smaller $\gamma_{\rm{res}}$ values and are thus more likely to exhibit intrinsic behavior.

\begin{figure}
\begin{center}
\includegraphics[width=\columnwidth,clip]{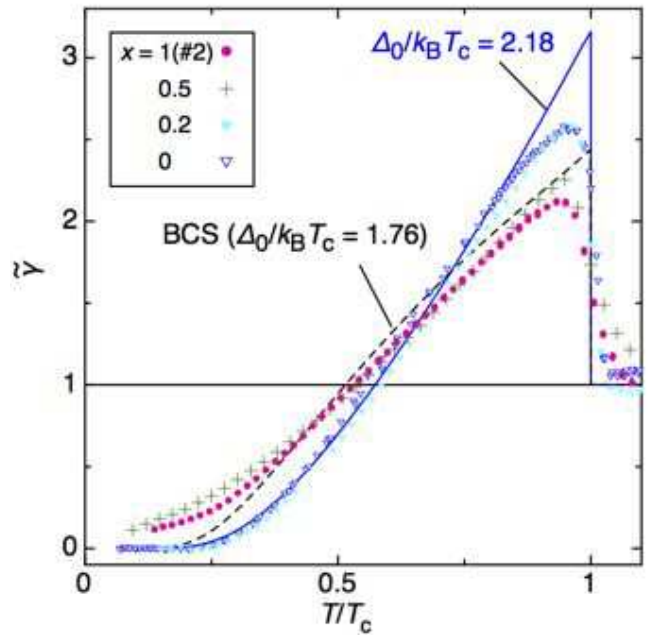}
\caption{(Color online)~Normalized superconducting electronic specific heat $\tilde{\gamma}  \equiv [c_{\rm{el}}(T)/T - \gamma_{\rm{res}}]/[\gamma_{\rm{n}} - \gamma_{\rm{res}}]$ for $x=0, 0.2, 0.5,$ and $1$($\#2$) plotted against the reduced temperature $T/T_{\rm{c}}$. The curves representing the conventional strong-coupling BCS model with $\Delta_0/k_{\rm{B}}T_{\rm{c}}=2.18$ and the weak-coupling BCS ($\Delta_0/k_{\rm{B}}T_{\rm{c}}=1.76$) model are also shown.
} 
\label{cel2}
\end{center}
\end{figure}

In order to examine the gap structure in more detail, we evaluate the normalized superconducting electronic specific heat: $\tilde{\gamma}  \equiv [c_{\rm{el}}(T)/T - \gamma_{\rm{res}}]/[\gamma_{\rm{n}} - \gamma_{\rm{res}}]$. We plot $\tilde{\gamma}$ of the $x=0, 0.2, 0.5,$ and $1$($\#2$) samples against the reduced temperature $T/T_{\rm{c}}$ in Fig.~\ref{cel2}(a). Curves based on the conventional strong-coupling BCS model~\cite{padamsee1973JLTP} with $\Delta_0/k_{\rm{B}}T_{\rm{c}}=2.18$ and the weak-coupling BCS model ($\Delta_0/k_{\rm{B}}T_{\rm{c}}=1.76$) are also presented. For $x\leq 0.2$, the temperature dependence of $\tilde{\gamma}$ is well explained by the former model. The $\Delta_0/k_{\rm{B}}T_{\rm{c}}$ value obtained is somewhat larger than that from another specific heat study~\cite{Takeya2007PRB} and the NMR $1/T_1$ result~\cite{Harada2012PRB}. A clearly reduced specific-heat jump at $T_{\rm{c}}$ is observed for $x=0.2$ and $x=0.5$. This change corresponds well with the change in the superconducting gap anisotropy revealed by the low-temperature behavior of $\tilde{\gamma}$. For $x=0.5$ and $x=1$($\#2$), the overall temperature dependence is similar; however, a noticeable deviation is seen at low temperatures. Note that the coherence peak in NMR $1/T_1$ just below $T_{\rm{c}}$ is observed for $x=0.5$, but is absent for $x=1$~\cite{Harada2012PRB}, indicating a spin-singlet-dominant state for $x=0.5$ and a spin-triplet-dominant state for $x=1$. Thus, the observed difference in $\tilde{\gamma}(T)$ is ascribable to this change in the spin singlet-triplet mixing ratio.

Here, we examine whether our results are explained by the simple model proposed by Yuan \textit{et al.}~\cite{Yuan2006PRL}. They assumed two Fermi surfaces split by the anisotropic spin-orbit interaction caused by the absence of inversion symmetry. They further assumed that the superconducting gap function of Li$_2$(Pd$_{1-x}$Pt$_x$)$_3$B belongs to the irreducible representation $A_1$, and the gap amplitude is expressed as $\Delta_{\pm}(\bm{k})=\Delta_1 \pm \Delta_2|\bm{g}(\bm{k})|$ for the two Fermi surfaces. Here, $\varDelta_1$ and $\varDelta_2$ are the magnitudes of the spin-singlet and spin-triplet components, respectively, and ${\bm{g}}({\bm{k}})$ is the dimensionless $g$-vector which represents the spin anisotropy due to anisotropic spin-orbit interaction. From the comparison with the penetration depth data, this model suggests that a fully-gapped spin-singlet-dominant state ($\Delta_1 > \Delta_2$) is realized for $x=0$, and a nodal spin-triplet-dominant state ($\Delta_1 < \Delta_2$) for $x=1$. In the spin-triplet-dominant state, the gap on one of the Fermi surfaces $\Delta_-(\bm{k})=\Delta_1-\Delta_2|\bm{g}(\bm{k})|$ has line nodes, whereas the other Fermi surface $\Delta_+(\bm{k})=\Delta_1+\Delta_2|\bm{g}(\bm{k})|$ remains fully-gapped as long as $\Delta_1 \neq 0$. 

The spin-singlet-dominant state, as evidenced by the existence of a coherence peak of the NMR $1/T_1$, has been reported for $x=0.5$~\cite{Harada2012PRB}. On the other hand, the emergence of an anisotropic superconducting gap is also evident even at $x=0.5$ by the power-law behavior of $\tilde{\gamma}$ in Fig.~\ref{cel2}(a) and Ref.~\cite{Takeya2007PRB}. Even in the region where the spin-singlet component is dominant, anisotropic superconducting gaps develop with an increase of $\Delta_2$, which can lead to gap anisotropy~\cite{Sigrist2006JMMM}. On further increasing $x$, this singlet-dominant anisotropic state changes to the triplet dominant nodal state above $x \sim 0.9$. This change is possibly triggered by the change in the local lattice distortion, which also causes the phonon anomaly revealed in this study. 

At this stage, the role of this phonon anomaly in the superconductivity is not clear. Further investigation of the relation between the phonon spectrum and the superconductivity as a function of $x$ is an interesting future issue.


In summary, we investigated the Pd-Pt substitution dependence of properties of the noncentrosymmetric superconductor Li$_2$(Pd$_{1-x}$Pt$_x$)$_3$B by specific heat. We find that the deviation of the phononic specific heat from the conventional Debye law grows as $x$ increases. The deviation exhibits a broad maximum around $x \sim 0.9$, indicating a phonon anomaly developing toward the lattice instability at $x \sim 0.9$~\cite{Harada2012PRB}. In the superconducting state, a power-law temperature dependence of the electronic specific heat for $x=0.5$ suggests the presence of anisotropy or nodes in the superconducting gap. 
Considering the correlation between the power-law temperature dependence and the residual coefficient, it is most probable that the anisotropic gap developing even at $x=0.5$, persists into the region beyond the lattice instability at $x \sim 0.9$, where a spin-triplet-dominant state is suggested.

We thank T. Shishidou, Y. Yanase, Y. Fuseya, C. Michioka, and K. Yoshimura for fruitful discussions and useful advice. This work was supported by a Grant-in-Aid for the Global COE program ``The Next Generation of Physics, Spun from Universality and Emergence'' from the Ministry of Education, Culture, Sports, Science, and Technology (MEXT) of Japan, and by the ``Topological Quantum Phenomena'' Grant-in Aid for Scientific Research on innovative Areas from MEXT of Japan. G.E., D.C.P., and M. K. were supported by the Japan Society for the Promotion of Science (JSPS).

\bibliographystyle{apsrev4-1_nocomma}
\bibliography{Preload,myrefs}

\begin{thebibliography}{22}%
\makeatletter
\providecommand \@ifxundefined [1]{%
 \@ifx{#1\undefined}
}%
\providecommand \@ifnum [1]{%
 \ifnum #1\expandafter \@firstoftwo
 \else \expandafter \@secondoftwo
 \fi
}%
\providecommand \@ifx [1]{%
 \ifx #1\expandafter \@firstoftwo
 \else \expandafter \@secondoftwo
 \fi
}%
\providecommand \natexlab [1]{#1}%
\providecommand \enquote  [1]{``#1''}%
\providecommand \bibnamefont  [1]{#1}%
\providecommand \bibfnamefont [1]{#1}%
\providecommand \citenamefont [1]{#1}%
\providecommand \href@noop [0]{\@secondoftwo}%
\providecommand \href [0]{\begingroup \@sanitize@url \@href}%
\providecommand \@href[1]{\@@startlink{#1}\@@href}%
\providecommand \@@href[1]{\endgroup#1\@@endlink}%
\providecommand \@sanitize@url [0]{\catcode `\\12\catcode `\$12\catcode
  `\&12\catcode `\#12\catcode `\^12\catcode `\_12\catcode `\%12\relax}%
\providecommand \@@startlink[1]{}%
\providecommand \@@endlink[0]{}%
\providecommand \url  [0]{\begingroup\@sanitize@url \@url }%
\providecommand \@url [1]{\endgroup\@href {#1}{\urlprefix }}%
\providecommand \urlprefix  [0]{URL }%
\providecommand \Eprint [0]{\href }%
\@ifxundefined \urlstyle {%
  \providecommand \doi  [0]{\begingroup \@sanitize@url \@doi}%
  \providecommand \@doi [1]{\endgroup \@@startlink {\doibase
  #1}doi:\discretionary {}{}{}#1\@@endlink }%
}{%
  \providecommand \doi  [0]{doi:\discretionary{}{}{}\begingroup
  \urlstyle{rm}\Url }%
}%
\providecommand \doibase [0]{http://dx.doi.org/}%
\providecommand \Doi [0]{\begingroup \@sanitize@url \@Doi }%
\providecommand \@Doi  [1]{\endgroup\@@startlink{\doibase#1}\@@Doi}%
\providecommand \@@Doi [1]{#1\@@endlink}%
\providecommand \selectlanguage [0]{\@gobble}%
\providecommand \bibinfo  [0]{\@secondoftwo}%
\providecommand \bibfield  [0]{\@secondoftwo}%
\providecommand \translation [1]{[#1]}%
\providecommand \BibitemOpen [0]{}%
\providecommand \bibitemStop [0]{}%
\providecommand \bibitemNoStop [0]{.\EOS\space}%
\providecommand \EOS [0]{\spacefactor3000\relax}%
\providecommand \BibitemShut  [1]{\csname bibitem#1\endcsname}%
\bibitem [{\citenamefont {Bauer}\ \emph {et~al.}(2004)\citenamefont {Bauer},
  \citenamefont {Hilscher}, \citenamefont {Michor}, \citenamefont {Paul},
  \citenamefont {Scheidt}, \citenamefont {Gribanov}, \citenamefont {Seropegin},
  \citenamefont {No{\"e}l}, \citenamefont {Sigrist},\ and\ \citenamefont
  {Rogl}}]{Bauer2004PRL}%
  \BibitemOpen
  \bibfield  {author} {\bibinfo {author} {\bibfnamefont {E.}~\bibnamefont
  {Bauer}}, \bibinfo {author} {\bibfnamefont {G.}~\bibnamefont {Hilscher}},
  \bibinfo {author} {\bibfnamefont {H.}~\bibnamefont {Michor}}, \bibinfo
  {author} {\bibfnamefont {C.}~\bibnamefont {Paul}}, \bibinfo {author}
  {\bibfnamefont {E.~W.}\ \bibnamefont {Scheidt}}, \bibinfo {author}
  {\bibfnamefont {A.}~\bibnamefont {Gribanov}}, \bibinfo {author}
  {\bibfnamefont {Y.}~\bibnamefont {Seropegin}}, \bibinfo {author}
  {\bibfnamefont {H.}~\bibnamefont {No{\"e}l}}, \bibinfo {author}
  {\bibfnamefont {M.}~\bibnamefont {Sigrist}}, \ and\ \bibinfo {author}
  {\bibfnamefont {P.}~\bibnamefont {Rogl}},\ }\Doi
  {10.1103/PhysRevLett.92.027003} {\bibfield  {journal} {\bibinfo  {journal}
  {Phys.\ Rev.\ Lett.}\ }\textbf {\bibinfo {volume} {92}},\ \bibinfo {pages}
  {027003} (\bibinfo {year} {2004})}\BibitemShut {NoStop}%
\bibitem [{\citenamefont {Kimura}\ \emph {et~al.}(2007)\citenamefont {Kimura},
  \citenamefont {Ito}, \citenamefont {Aoki}, \citenamefont {Uji},\ and\
  \citenamefont {Terashima}}]{Kimura2007PRL}%
  \BibitemOpen
  \bibfield  {author} {\bibinfo {author} {\bibfnamefont {N.}~\bibnamefont
  {Kimura}}, \bibinfo {author} {\bibfnamefont {K.}~\bibnamefont {Ito}},
  \bibinfo {author} {\bibfnamefont {H.}~\bibnamefont {Aoki}}, \bibinfo {author}
  {\bibfnamefont {S.}~\bibnamefont {Uji}}, \ and\ \bibinfo {author}
  {\bibfnamefont {T.}~\bibnamefont {Terashima}},\ }\Doi
  {10.1103/PhysRevLett.98.197001} {\bibfield  {journal} {\bibinfo  {journal}
  {Phys. Rev. Lett.}\ }\textbf {\bibinfo {volume} {98}},\ \bibinfo {pages}
  {197001} (\bibinfo {year} {2007})}\BibitemShut {NoStop}%
\bibitem [{\citenamefont {Sigrist}\ \emph {et~al.}(2007)\citenamefont
  {Sigrist}, \citenamefont {Agterberg}, \citenamefont {Frigeri}, \citenamefont
  {Hayashi}, \citenamefont {Kaur}, \citenamefont {Koga}, \citenamefont {Milat},
  \citenamefont {Wakabayashi},\ and\ \citenamefont {Yanase}}]{Sigrist2006JMMM}%
  \BibitemOpen
  \bibfield  {author} {\bibinfo {author} {\bibfnamefont {M.}~\bibnamefont
  {Sigrist}}, \bibinfo {author} {\bibfnamefont {D.}~\bibnamefont {Agterberg}},
  \bibinfo {author} {\bibfnamefont {P.}~\bibnamefont {Frigeri}}, \bibinfo
  {author} {\bibfnamefont {N.}~\bibnamefont {Hayashi}}, \bibinfo {author}
  {\bibfnamefont {R.}~\bibnamefont {Kaur}}, \bibinfo {author} {\bibfnamefont
  {A.}~\bibnamefont {Koga}}, \bibinfo {author} {\bibfnamefont {I.}~\bibnamefont
  {Milat}}, \bibinfo {author} {\bibfnamefont {K.}~\bibnamefont {Wakabayashi}},
  \ and\ \bibinfo {author} {\bibfnamefont {Y.}~\bibnamefont {Yanase}},\ }\Doi
  {10.1016/j.jmmm.2006.10.141} {\bibfield  {journal} {\bibinfo  {journal} {J.\
  Magn.\ Magn.\ Mat.}\ }\textbf {\bibinfo {volume} {310}},\ \bibinfo {pages}
  {536} (\bibinfo {year} {2007})}\BibitemShut {NoStop}%
\bibitem [{\citenamefont {Agterberg}\ and\ \citenamefont
  {Kaur}(2007)}]{Agterberg2007PRB}%
  \BibitemOpen
  \bibfield  {author} {\bibinfo {author} {\bibfnamefont {D.~F.}\ \bibnamefont
  {Agterberg}}\ and\ \bibinfo {author} {\bibfnamefont {R.~P.}\ \bibnamefont
  {Kaur}},\ }\Doi {10.1103/PhysRevB.75.064511} {\bibfield  {journal} {\bibinfo
  {journal} {Phys.\ Rev.\ B}\ }\textbf {\bibinfo {volume} {75}},\ \bibinfo
  {pages} {064511} (\bibinfo {year} {2007})}\BibitemShut {NoStop}%
\bibitem [{\citenamefont {Fu}\ and\ \citenamefont {Kane}(2008)}]{Fu2008PRL}%
  \BibitemOpen
  \bibfield  {author} {\bibinfo {author} {\bibfnamefont {L.}~\bibnamefont
  {Fu}}\ and\ \bibinfo {author} {\bibfnamefont {C.~L.}\ \bibnamefont {Kane}},\
  }\Doi {10.1103/PhysRevLett.100.096407} {\bibfield  {journal} {\bibinfo
  {journal} {Phys. Rev. Lett.}\ }\textbf {\bibinfo {volume} {100}},\ \bibinfo
  {pages} {096407} (\bibinfo {year} {2008})}\BibitemShut {NoStop}%
\bibitem [{\citenamefont {Sato}\ \emph {et~al.}(2009)\citenamefont {Sato},
  \citenamefont {Takahashi},\ and\ \citenamefont {Fujimoto}}]{Sato2009PRL}%
  \BibitemOpen
  \bibfield  {author} {\bibinfo {author} {\bibfnamefont {M.}~\bibnamefont
  {Sato}}, \bibinfo {author} {\bibfnamefont {Y.}~\bibnamefont {Takahashi}}, \
  and\ \bibinfo {author} {\bibfnamefont {S.}~\bibnamefont {Fujimoto}},\ }\Doi
  {10.1103/PhysRevLett.103.020401} {\bibfield  {journal} {\bibinfo  {journal}
  {Phys. Rev. Lett.}\ }\textbf {\bibinfo {volume} {103}},\ \bibinfo {pages}
  {020401} (\bibinfo {year} {2009})}\BibitemShut {NoStop}%
\bibitem [{\citenamefont {Sato}\ \emph {et~al.}(2010)\citenamefont {Sato},
  \citenamefont {Takahashi},\ and\ \citenamefont {Fujimoto}}]{Sato2010PRB}%
  \BibitemOpen
  \bibfield  {author} {\bibinfo {author} {\bibfnamefont {M.}~\bibnamefont
  {Sato}}, \bibinfo {author} {\bibfnamefont {Y.}~\bibnamefont {Takahashi}}, \
  and\ \bibinfo {author} {\bibfnamefont {S.}~\bibnamefont {Fujimoto}},\ }\Doi
  {10.1103/PhysRevB.82.134521} {\bibfield  {journal} {\bibinfo  {journal}
  {Phys. Rev. B}\ }\textbf {\bibinfo {volume} {82}},\ \bibinfo {pages} {134521}
  (\bibinfo {year} {2010})}\BibitemShut {NoStop}%
\bibitem [{\citenamefont {Yuan}\ \emph {et~al.}(2006)\citenamefont {Yuan},
  \citenamefont {Agterberg}, \citenamefont {Hayashi}, \citenamefont {Badica},
  \citenamefont {Vandervelde}, \citenamefont {Togano}, \citenamefont
  {Sigrist},\ and\ \citenamefont {Salamon}}]{Yuan2006PRL}%
  \BibitemOpen
  \bibfield  {author} {\bibinfo {author} {\bibfnamefont {H.~Q.}\ \bibnamefont
  {Yuan}}, \bibinfo {author} {\bibfnamefont {D.~F.}\ \bibnamefont {Agterberg}},
  \bibinfo {author} {\bibfnamefont {N.}~\bibnamefont {Hayashi}}, \bibinfo
  {author} {\bibfnamefont {P.}~\bibnamefont {Badica}}, \bibinfo {author}
  {\bibfnamefont {D.}~\bibnamefont {Vandervelde}}, \bibinfo {author}
  {\bibfnamefont {K.}~\bibnamefont {Togano}}, \bibinfo {author} {\bibfnamefont
  {M.}~\bibnamefont {Sigrist}}, \ and\ \bibinfo {author} {\bibfnamefont
  {M.~B.}\ \bibnamefont {Salamon}},\ }\Doi {10.1103/PhysRevLett.97.017006}
  {\bibfield  {journal} {\bibinfo  {journal} {Phys. Rev. Lett.}\ }\textbf
  {\bibinfo {volume} {97}},\ \bibinfo {pages} {017006} (\bibinfo {year}
  {2006})}\BibitemShut {NoStop}%
\bibitem [{\citenamefont {Nishiyama}\ \emph {et~al.}(2007)\citenamefont
  {Nishiyama}, \citenamefont {Inada},\ and\ \citenamefont
  {Zheng}}]{Nishiyama2007PRL}%
  \BibitemOpen
  \bibfield  {author} {\bibinfo {author} {\bibfnamefont {M.}~\bibnamefont
  {Nishiyama}}, \bibinfo {author} {\bibfnamefont {Y.}~\bibnamefont {Inada}}, \
  and\ \bibinfo {author} {\bibfnamefont {G.-q.}\ \bibnamefont {Zheng}},\ }\Doi
  {10.1103/PhysRevLett.98.047002} {\bibfield  {journal} {\bibinfo  {journal}
  {Phys. Rev. Lett.}\ }\textbf {\bibinfo {volume} {98}},\ \bibinfo {pages}
  {047002} (\bibinfo {year} {2007})}\BibitemShut {NoStop}%
\bibitem [{\citenamefont {Harada}\ \emph {et~al.}(2012)\citenamefont {Harada},
  \citenamefont {Zhou}, \citenamefont {Yao}, \citenamefont {Inada},\ and\
  \citenamefont {Zheng}}]{Harada2012PRB}%
  \BibitemOpen
  \bibfield  {author} {\bibinfo {author} {\bibfnamefont {S.}~\bibnamefont
  {Harada}}, \bibinfo {author} {\bibfnamefont {J.~J.}\ \bibnamefont {Zhou}},
  \bibinfo {author} {\bibfnamefont {Y.~G.}\ \bibnamefont {Yao}}, \bibinfo
  {author} {\bibfnamefont {Y.}~\bibnamefont {Inada}}, \ and\ \bibinfo {author}
  {\bibfnamefont {G.-q.}\ \bibnamefont {Zheng}},\ }\Doi
  {10.1103/PhysRevB.86.220502} {\bibfield  {journal} {\bibinfo  {journal}
  {Phys. Rev. B}\ }\textbf {\bibinfo {volume} {86}},\ \bibinfo {pages} {220502}
  (\bibinfo {year} {2012})}\BibitemShut {NoStop}%
\bibitem [{\citenamefont {Lee}\ and\ \citenamefont
  {Pickett}(2005)}]{Lee2005PRB}%
  \BibitemOpen
  \bibfield  {author} {\bibinfo {author} {\bibfnamefont {K.-W.}\ \bibnamefont
  {Lee}}\ and\ \bibinfo {author} {\bibfnamefont {W.~E.}\ \bibnamefont
  {Pickett}},\ }\Doi {10.1103/PhysRevB.72.174505} {\bibfield  {journal}
  {\bibinfo  {journal} {Phys. Rev. B}\ }\textbf {\bibinfo {volume} {72}},\
  \bibinfo {pages} {174505} (\bibinfo {year} {2005})}\BibitemShut {NoStop}%
\bibitem [{\citenamefont {Chandra}\ \emph {et~al.}(2005)\citenamefont
  {Chandra}, \citenamefont {Jaya},\ and\ \citenamefont
  {Valsakumar}}]{Chandra2005116}%
  \BibitemOpen
  \bibfield  {author} {\bibinfo {author} {\bibfnamefont {S.}~\bibnamefont
  {Chandra}}, \bibinfo {author} {\bibfnamefont {S.~M.}\ \bibnamefont {Jaya}}, \
  and\ \bibinfo {author} {\bibfnamefont {M.}~\bibnamefont {Valsakumar}},\ }\Doi
  {10.1016/j.physc.2005.07.018} {\bibfield  {journal} {\bibinfo  {journal}
  {Physica C: Superconductivity}\ }\textbf {\bibinfo {volume} {432}},\ \bibinfo
  {pages} {116 } (\bibinfo {year} {2005})}\BibitemShut {NoStop}%
\bibitem [{\citenamefont {Yoshida}\ \emph {et~al.}(2008)\citenamefont
  {Yoshida}, \citenamefont {Okazaki}, \citenamefont {Tajima}, \citenamefont
  {Muro}, \citenamefont {Hase}, \citenamefont {Okada}, \citenamefont {Takeya},
  \citenamefont {Hirata}, \citenamefont {Hirai}, \citenamefont {Muraoka},\ and\
  \citenamefont {Yokoya}}]{Yoshida2008JPSJ}%
  \BibitemOpen
  \bibfield  {author} {\bibinfo {author} {\bibfnamefont {R.}~\bibnamefont
  {Yoshida}}, \bibinfo {author} {\bibfnamefont {H.}~\bibnamefont {Okazaki}},
  \bibinfo {author} {\bibfnamefont {M.}~\bibnamefont {Tajima}}, \bibinfo
  {author} {\bibfnamefont {T.}~\bibnamefont {Muro}}, \bibinfo {author}
  {\bibfnamefont {I.}~\bibnamefont {Hase}}, \bibinfo {author} {\bibfnamefont
  {K.}~\bibnamefont {Okada}}, \bibinfo {author} {\bibfnamefont
  {H.}~\bibnamefont {Takeya}}, \bibinfo {author} {\bibfnamefont
  {K.}~\bibnamefont {Hirata}}, \bibinfo {author} {\bibfnamefont
  {M.}~\bibnamefont {Hirai}}, \bibinfo {author} {\bibfnamefont
  {Y.}~\bibnamefont {Muraoka}}, \ and\ \bibinfo {author} {\bibfnamefont
  {T.}~\bibnamefont {Yokoya}},\ }\Doi {10.1143/JPSJ.77.104701} {\bibfield
  {journal} {\bibinfo  {journal} {J.\ Phys.\ Soc.\ Jpn.}\ }\textbf {\bibinfo
  {volume} {77}},\ \bibinfo {pages} {104701} (\bibinfo {year}
  {2008})}\BibitemShut {NoStop}%
\bibitem [{\citenamefont {Tsuda}\ \emph {et~al.}(2009)\citenamefont {Tsuda},
  \citenamefont {Yokoya}, \citenamefont {Kiss}, \citenamefont {Shimojima},
  \citenamefont {Ishizaka}, \citenamefont {Shin}, \citenamefont {Togashi},
  \citenamefont {Watanabe}, \citenamefont {Zhang}, \citenamefont {Chen},
  \citenamefont {Hase}, \citenamefont {Takeya}, \citenamefont {Hirata},\ and\
  \citenamefont {Togano}}]{Tsuda2009JPSJ}%
  \BibitemOpen
  \bibfield  {author} {\bibinfo {author} {\bibfnamefont {S.}~\bibnamefont
  {Tsuda}}, \bibinfo {author} {\bibfnamefont {T.}~\bibnamefont {Yokoya}},
  \bibinfo {author} {\bibfnamefont {T.}~\bibnamefont {Kiss}}, \bibinfo {author}
  {\bibfnamefont {T.}~\bibnamefont {Shimojima}}, \bibinfo {author}
  {\bibfnamefont {K.}~\bibnamefont {Ishizaka}}, \bibinfo {author}
  {\bibfnamefont {S.}~\bibnamefont {Shin}}, \bibinfo {author} {\bibfnamefont
  {T.}~\bibnamefont {Togashi}}, \bibinfo {author} {\bibfnamefont
  {S.}~\bibnamefont {Watanabe}}, \bibinfo {author} {\bibfnamefont
  {C.}~\bibnamefont {Zhang}}, \bibinfo {author} {\bibfnamefont
  {C.}~\bibnamefont {Chen}}, \bibinfo {author} {\bibfnamefont {I.}~\bibnamefont
  {Hase}}, \bibinfo {author} {\bibfnamefont {H.}~\bibnamefont {Takeya}},
  \bibinfo {author} {\bibfnamefont {K.}~\bibnamefont {Hirata}}, \ and\ \bibinfo
  {author} {\bibfnamefont {K.}~\bibnamefont {Togano}},\ }\Doi
  {10.1143/JPSJ.78.034711} {\bibfield  {journal} {\bibinfo  {journal} {J.\
  Phys.\ Soc.\ Jpn.}\ }\textbf {\bibinfo {volume} {78}},\ \bibinfo {pages}
  {034711} (\bibinfo {year} {2009})}\BibitemShut {NoStop}%
\bibitem [{\citenamefont {Takeya}\ \emph {et~al.}(2007)\citenamefont {Takeya},
  \citenamefont {ElMassalami}, \citenamefont {Kasahara},\ and\ \citenamefont
  {Hirata}}]{Takeya2007PRB}%
  \BibitemOpen
  \bibfield  {author} {\bibinfo {author} {\bibfnamefont {H.}~\bibnamefont
  {Takeya}}, \bibinfo {author} {\bibfnamefont {M.}~\bibnamefont {ElMassalami}},
  \bibinfo {author} {\bibfnamefont {S.}~\bibnamefont {Kasahara}}, \ and\
  \bibinfo {author} {\bibfnamefont {K.}~\bibnamefont {Hirata}},\ }\Doi
  {10.1103/PhysRevB.76.104506} {\bibfield  {journal} {\bibinfo  {journal}
  {Phys. Rev. B}\ }\textbf {\bibinfo {volume} {76}},\ \bibinfo {pages} {104506}
  (\bibinfo {year} {2007})}\BibitemShut {NoStop}%
\bibitem [{\citenamefont {H{\"a}fliger}\ \emph {et~al.}(2009)\citenamefont
  {H{\"a}fliger}, \citenamefont {Khasanov}, \citenamefont {Lortz},
  \citenamefont {Petrovi{\'c}}, \citenamefont {Togano}, \citenamefont {Baines},
  \citenamefont {Graneli},\ and\ \citenamefont {Keller}}]{Hafliger2009JSNM}%
  \BibitemOpen
  \bibfield  {author} {\bibinfo {author} {\bibfnamefont {P.}~\bibnamefont
  {H{\"a}fliger}}, \bibinfo {author} {\bibfnamefont {R.}~\bibnamefont
  {Khasanov}}, \bibinfo {author} {\bibfnamefont {R.}~\bibnamefont {Lortz}},
  \bibinfo {author} {\bibfnamefont {A.}~\bibnamefont {Petrovi{\'c}}}, \bibinfo
  {author} {\bibfnamefont {K.}~\bibnamefont {Togano}}, \bibinfo {author}
  {\bibfnamefont {C.}~\bibnamefont {Baines}}, \bibinfo {author} {\bibfnamefont
  {B.}~\bibnamefont {Graneli}}, \ and\ \bibinfo {author} {\bibfnamefont
  {H.}~\bibnamefont {Keller}},\ }\href@noop {} {\bibfield  {journal} {\bibinfo
  {journal} {J.\ Supercond.\ Nov.\ Magn.}\ }\textbf {\bibinfo {volume} {22}},\
  \bibinfo {pages} {337} (\bibinfo {year} {2009})}\BibitemShut {NoStop}%
\bibitem [{\citenamefont {Peets}\ \emph {et~al.}(2011)\citenamefont {Peets},
  \citenamefont {Eguchi}, \citenamefont {Kriener}, \citenamefont {Harada},
  \citenamefont {Shamsuzzamen}, \citenamefont {Inada}, \citenamefont {Zheng},\
  and\ \citenamefont {Maeno}}]{Darren2011PRB}%
  \BibitemOpen
  \bibfield  {author} {\bibinfo {author} {\bibfnamefont {D.~C.}\ \bibnamefont
  {Peets}}, \bibinfo {author} {\bibfnamefont {G.}~\bibnamefont {Eguchi}},
  \bibinfo {author} {\bibfnamefont {M.}~\bibnamefont {Kriener}}, \bibinfo
  {author} {\bibfnamefont {S.}~\bibnamefont {Harada}}, \bibinfo {author}
  {\bibfnamefont {S.~M.}\ \bibnamefont {Shamsuzzamen}}, \bibinfo {author}
  {\bibfnamefont {Y.}~\bibnamefont {Inada}}, \bibinfo {author} {\bibfnamefont
  {G.-q.}\ \bibnamefont {Zheng}}, \ and\ \bibinfo {author} {\bibfnamefont
  {Y.}~\bibnamefont {Maeno}},\ }\Doi {10.1103/PhysRevB.84.054521} {\bibfield
  {journal} {\bibinfo  {journal} {Phys. Rev. B}\ }\textbf {\bibinfo {volume}
  {84}},\ \bibinfo {pages} {054521} (\bibinfo {year} {2011})}\BibitemShut
  {NoStop}%
\bibitem [{\citenamefont {Badica}\ \emph {et~al.}(2005)\citenamefont {Badica},
  \citenamefont {Kondo},\ and\ \citenamefont {Togano}}]{Badica2005JPSJ}%
  \BibitemOpen
  \bibfield  {author} {\bibinfo {author} {\bibfnamefont {P.}~\bibnamefont
  {Badica}}, \bibinfo {author} {\bibfnamefont {T.}~\bibnamefont {Kondo}}, \
  and\ \bibinfo {author} {\bibfnamefont {K.}~\bibnamefont {Togano}},\
  }\href@noop {} {\bibfield  {journal} {\bibinfo  {journal} {J.\ Phys.\ Soc.\
  Jpn.}\ }\textbf {\bibinfo {volume} {74}},\ \bibinfo {pages} {pp. 1014}
  (\bibinfo {year} {2005})}\BibitemShut {NoStop}%
\bibitem [{\citenamefont {Harada}\ \emph {et~al.}(2010)\citenamefont {Harada},
  \citenamefont {Inada},\ and\ \citenamefont {Zheng}}]{Harada2010PhysC}%
  \BibitemOpen
  \bibfield  {author} {\bibinfo {author} {\bibfnamefont {S.}~\bibnamefont
  {Harada}}, \bibinfo {author} {\bibfnamefont {Y.}~\bibnamefont {Inada}}, \
  and\ \bibinfo {author} {\bibfnamefont {G.-q.}\ \bibnamefont {Zheng}},\ }\Doi
  {10.1016/j.physc.2010.05.043} {\bibfield  {journal} {\bibinfo  {journal}
  {Physica C}\ }\textbf {\bibinfo {volume} {470}},\ \bibinfo {pages} {1089}
  (\bibinfo {year} {2010})}\BibitemShut {NoStop}%
\bibitem [{\citenamefont {Shamsuzzamen}\ \emph {et~al.}(2010)\citenamefont
  {Shamsuzzamen}, \citenamefont {Inada}, \citenamefont {Sasano}, \citenamefont
  {Harada},\ and\ \citenamefont {Zheng}}]{Shamsuzzamen2010JPCO}%
  \BibitemOpen
  \bibfield  {author} {\bibinfo {author} {\bibfnamefont {S.~M.}\ \bibnamefont
  {Shamsuzzamen}}, \bibinfo {author} {\bibfnamefont {Y.}~\bibnamefont {Inada}},
  \bibinfo {author} {\bibfnamefont {S.}~\bibnamefont {Sasano}}, \bibinfo
  {author} {\bibfnamefont {S.}~\bibnamefont {Harada}}, \ and\ \bibinfo {author}
  {\bibfnamefont {G.-q.}\ \bibnamefont {Zheng}},\ }\Doi
  {10.1088/1742-6596/200/1/012183} {\bibfield  {journal} {\bibinfo  {journal}
  {J.\ Phys.: Conf.\ Ser.}\ }\textbf {\bibinfo {volume} {200}},\ \bibinfo
  {pages} {012183} (\bibinfo {year} {2010})}\BibitemShut {NoStop}%
\bibitem [{\citenamefont {Chang}\ \emph {et~al.}(2000)\citenamefont {Chang},
  \citenamefont {Lin},\ and\ \citenamefont {Yang}}]{Chang2000PRB}%
  \BibitemOpen
  \bibfield  {author} {\bibinfo {author} {\bibfnamefont {C.~F.}\ \bibnamefont
  {Chang}}, \bibinfo {author} {\bibfnamefont {J.-Y.}\ \bibnamefont {Lin}}, \
  and\ \bibinfo {author} {\bibfnamefont {H.~D.}\ \bibnamefont {Yang}},\ }\Doi
  {10.1103/PhysRevB.61.14350} {\bibfield  {journal} {\bibinfo  {journal} {Phys.
  Rev. B}\ }\textbf {\bibinfo {volume} {61}},\ \bibinfo {pages} {14350}
  (\bibinfo {year} {2000})}\BibitemShut {NoStop}%
\bibitem [{\citenamefont {Padamsee}\ \emph {et~al.}(1973)\citenamefont
  {Padamsee}, \citenamefont {Neighbor},\ and\ \citenamefont
  {Shiffman}}]{padamsee1973JLTP}%
  \BibitemOpen
  \bibfield  {author} {\bibinfo {author} {\bibfnamefont {H.}~\bibnamefont
  {Padamsee}}, \bibinfo {author} {\bibfnamefont {J.}~\bibnamefont {Neighbor}},
  \ and\ \bibinfo {author} {\bibfnamefont {C.}~\bibnamefont {Shiffman}},\
  }\href@noop {} {\bibfield  {journal} {\bibinfo  {journal} {J.\ Low Temp.\
  Phys.}\ }\textbf {\bibinfo {volume} {12}},\ \bibinfo {pages} {387} (\bibinfo
  {year} {1973})}\BibitemShut {NoStop}%
\end{thebibliography}%

\end{document}